\documentclass[12pt,letterpaper]{article}

\usepackage{amsmath}
\usepackage[psamsfonts]{amssymb}
\usepackage{amsthm}
\usepackage{fullpage}
\usepackage{cite}
\usepackage{indentfirst}
\usepackage{graphicx}
\usepackage{xspace}

\newcommand{\R}{{\mathbf R}}
\newcommand{\C}{{\mathbf C}}
\newcommand{\PP}{{\mathbf P}}
\newcommand{\N}{{\mathcal N}}

\makeatletter
\@addtoreset{equation}{section}
\renewcommand\theequation{\thesection.\@arabic\c@equation}
\makeatother

\newcommand{\CPP}{\C\PP^{3|4}}
\newcommand{\RPP}{\R\PP^{3|4}}

\newcommand{\Tr}{{\mathrm{Tr}}\xspace}

\begin{document}

\bibliographystyle{utphys}

\setcounter{page}{1}
\pagestyle{plain}

\begin{titlepage}

\begin{center}
\hfill HUTP-04/A005\\
\hfill hep-th/0402128

\vskip 1.5 cm
{\huge \bf $\N=2$ strings and the twistorial Calabi-Yau}
\vskip 1.3 cm
{\large Andrew Neitzke and Cumrun Vafa}\\
\vskip 0.5 cm
{Jefferson Physical Laboratory,
Harvard University,\\
Cambridge, MA 02138, USA}
\vskip 0.3cm
{neitzke@fas.harvard.edu\\
vafa@physics.harvard.edu}
\end{center}

\vskip 0.5 cm
\begin{abstract}
We interpret the A and B model topological strings
on $\CPP$ as equivalent to open $\N=2$ string theory
on spacetime with signature $(2,2)$, when covariantized
with respect to $SO(2,2)$ and supersymmetrized a la Siegel.
We propose that instantons 
ending on Lagrangian branes wrapping $\RPP$
deform the self-dual $\N=4$ Yang-Mills sector to
ordinary Yang-Mills by generating
a `t Hooft like expansion.  We conjecture that the A and
B versions are S-dual to each other.  We also conjecture
that mirror symmetry may explain the recent observations
of Witten that twistor transformed $\N=4$ Yang-Mills
amplitudes lie on holomorphic curves.
\end{abstract}

\end{titlepage}

\renewcommand{\baselinestretch}{1.4}
\small\normalsize

\section{Introduction}

In a beautiful recent paper \cite{Witten:2003nn}, 
Witten provided highly non-trivial evidence
that the planar amplitudes of $\N=4$ supersymmetric $U(N)$
Yang-Mills theory in four 
dimensions, once transformed 
to the twistor space $\CPP$, are supported on holomorphic curves (see 
also the followup work \cite{Roiban:2004vt,Berkovits:2004hg,Roiban:2004ka}.)
This work was based on the spinor formalism for gauge theory amplitudes;
see e.g. the early work \cite{Parke:1985pn,Parke:1986gb,Bern:1993mq} 
and more recent papers \cite{Bern:2002tk,Bern:2003ck}.
Since holomorphic curves are traditional hallmarks of string theory,
it was attempted in \cite{Witten:2003nn} to provide a tentative string theory interpretation of 
their appearance.  This interpretation involved 
the topological B model on the Calabi-Yau manifold
$\CPP$, with $N$ D5-branes wrapping the Calabi-Yau.
The computation of the amplitude then reduced to a current algebra calculation
on the holomorphic curves as suggested long ago in \cite{Nair:1988bq}.
However, the objects lying on the holomorphic curves were 
proposed to be closed D1-instantons.

There were a number of puzzles associated
with these D1-instantons:  i) The amplitude
came out as a holomorphic $n$-form on an instanton moduli 
space of complex dimension $n$; such a form can
be integrated only after choosing a contour, but
it was not clear a priori which contour to choose;
ii) Only the planar gauge theory amplitudes were considered; it 
was not clear how non-planar contributions could fit in
(by this we mean `t Hooft diagrams 
with handles \cite{'tHooft:1974jz}).

It could be that somehow a natural real cycle in the instanton moduli space
is identified, or alternatively that all nontrivial choices
are somehow equivalent, thus solving problem i).  As for problem ii), it could
be that this story is just a planar story, and the non-planar
diagrams just do not work in such a simple way.  In particular,
it could be that the non-planar diagrams cannot be decoupled
from the gravity sector and consistency of the theory
would require a fixed finite rank gauge group (a 
possibility we will briefly discuss
in the context of orientifolding the string realization,
although it seems unlikely since there is no sign of
any pathology for arbitrary rank $\N=4$ gauge theory on its own.)
However, it would be nice if one 
could find an alternative mechanism to solve these two puzzles, which 
would enhance the usefulness of the
twistor transform for arbitrary rank classical
gauge groups beyond the planar limit.
The main content of this 
paper is an exploration of one possible mechanism.  We will see that this possibility
leads to some intriguing ideas which, while far from proven, could merit
further investigation.  Moreover, some of the
ideas we encounter may be relevant even if the 
original puzzle is solved in other ways.

Here is the mechanism.  If somehow the D1-instantons were required to end on the
Lagrangian submanifold $\RPP \subset \CPP$, then the dimension 
of the moduli space would naturally be reduced by half.
Moreover, this could also solve the second problem; the
topology of the instantons would be characterized by
two parameters instead of one (the number of boundaries on 
$\RPP$ and the number of handles), so one could read off a `t Hooft-like
expansion.  For this `t Hooft counting to work out correctly,
we would need to have $N$ branes wrapping $\RPP$.  However,
this raises a puzzle:  What is the meaning
of these Lagrangian branes in the B model?  Certainly
they cannot be D-branes as that would violate the BRST
symmetry of the B model.

The need for Lagrangian branes would suggest an A model interpretation,
where we have $N$ Lagrangian D-branes wrapping on
$\RPP$ and the D1-brane instantons are replaced by worldsheet
instantons.  However, there would be a dual puzzle in this context
\cite{Witten:2003nn}:  There are too few states on the Lagrangian
D-brane to support the fields of $\N=4$ Yang-Mills\footnote{
We would like to thank E. Witten and H. Ooguri for discussions
on this point.}.  The moduli space of flat connections on $\RPP$ seems
to support at most a finite number of states and no continuous fields.

Here we suggest a tentative resolution to these puzzles, starting
in the B model.  We propose
that there could be a gravitational tadpole in the B model
which needs to be cancelled, and that this cancellation can
be accomplished by the introduction of $N$ 
Lagrangian branes (not D-branes) wrapped
on $\RPP$.  Moreover, if we assume the
existence of these Lagrangian branes (which
we call ``NS2-branes'') with their natural couplings, 
the D1-branes can end on them.   We interpret
these open D1-instantons as the relevant ones for deforming
self-dual Yang-Mills to the full Yang-Mills, following the picture
of \cite{Witten:2003nn}.  This would resolve the 
issue in the B model context. 

This B model story is quite similar to the A model picture which
we first considered above, and we make the tentative suggestion that 
the two are in fact related by S-duality.  
In other words, we propose that the Montonen-Olive duality at
the level of topological strings converts the B model into
an A model on the same Calabi-Yau.  In this language we can see a possible resolution 
to the problem of the A model not having enough physical states:  we
now have an extra ingredient, namely the $\CPP$ filling branes
which are S-dual to the D5-branes of the B model.  Borrowing
the terminology of type IIB superstrings, we will call these ``NS5-branes''.

The fact that $\RPP$ features prominently here
as the locus where instantons can end suggests
that this is the natural arena for the comparison to gauge theory.
In other words, we rephrase
the observations of \cite{Witten:2003nn} as the statement
that the amplitudes of $\N=4$ supersymmetric Yang-Mills
on spacetime with signature $++--$, when
twistor-transformed to functions
on $\RPP$, are supported on the real {\it boundaries} of holomorphic
curves in twistor space $\CPP$.  Then
the holomorphic curves in question would be identified as the boundaries of
instantons ending on the Lagrangian branes (either D1-instantons or
worldsheet instantons depending on whether we use the A or B model.)
In fact, it was already noted in \cite{Witten:2003nn} that
$\RPP$ can arise naturally if 
one specializes to the signature $++--$.  This signature 
is also natural from the viewpoint of $\N=2$ strings
\cite{Ooguri:1991fp}, and
as we will explain in the next section, we believe that
the topological theory in twistor space $\CPP$ is indeed
equivalent to $\N=2$ strings.

It is also natural to ask what role mirror 
symmetry can play in this story.  We have proposed
that S-duality exchanges the A and B models on the same
twistorial Calabi-Yau.  On the other hand, mirror
symmetry also exchanges the A and B model, but
also changes the Calabi-Yau in question.  We propose
that mirror symmetry, together with this S-duality,
may explain from first principles why the amplitudes of 
$\N=4$ Yang-Mills are localized on boundaries of holomorphic
curves on twistor space.  Mirror symmetry would map
the worldsheet instantons of the A model to field theory
computations of a B model on a different
Calabi-Yau, which we conjecture to be the
quadric on ${\bf CP}^{3|3}\times {\bf CP}^{3|3}$.  The
reason for this conjecture is simply that this space
is already known to lead to a twistorial formulation
of the full $\N=4$ Yang-Mills theory including
all interactions, at least at the level of 
the classical equations of motion, without requiring 
any instanton corrections \cite{Witten:1978xx}, as was
re-emphasized in \cite{Witten:2003nn}.

We have suggested that the natural signature for the $\N=4$ 
super Yang-Mills theory in this twistor correspondence is
the $++--$ signature, which is precisely
the one arising naturally for $\N=2$ strings.  
Now what about the gravitational sector?
The gravity version of this
theory has been suggested in \cite{Witten:2003nn} to
be $\N=4$ conformal supergravity.  This can in principle
be reconciled with an $\N=2$ closed string theory, with signature $++--$.
This would be viewed as the instanton corrected version of
$\N=4$ self-dual conformal gravity \cite{Ketov:1993ix}.
This also explains the fact that the propagators of the gravity
theory for $\N=2$ strings are expected to be $1/(k^2)^2$ \cite{Ooguri:1991fp}
as is the case for conformal gravity.
In this context it is amusing to note that a non-perturbative
definition of the target space gravity theory of
the topological A model has been proposed in \cite{Okounkov:2003sp,Iqbal:2003ds};
there the target space theory was described as a 
gravitational quantum foam.
This would suggest a gravitational quantum foam picture
in the A model on $\CPP$; the body of
this space, $\C\PP^3$, projects to the tetrahedral crystal.
It would be very interesting to clarify the nature
of this gravitational quantum foam for this
topological theory with the additional NS5-branes, as it would correspond
to the twistor transform of ordinary gravitational foam in
$\N=4$ conformal supergravity in four dimensions.

The organization of this paper is as follows.
At the end of this section we summarize how $\N=2$
strings enter the picture, as the relevant literature seems to be
unfamiliar to most readers, and is one of the main
motivations of this paper.  In Section 2 we briefly review 
topological A and B model strings with D-branes included.  
In Section 3 we argue for the existence of extra
branes in the A and B models.  In section
4 we show how these additional branes affect the
twistorial Calabi-Yau and we propose a role
for S-duality.  In section 5 we discuss
the possible role of mirror symmetry in this story.  In section 6
we discuss some additional issues.

\subsection{Historical background}

The string worldsheet can be viewed as a theory of 2d gravity
coupled to matter.  The corresponding 2d gravity theories can
be classified by the number of supersymmetries $\N$.  Bosonic strings
and superstrings correspond to $\N=0$ and $\N=1$ worldsheet 
supersymmetry respectively, and the heterotic
string is where left-movers and right-movers have $(\N_L,\N_R)=(0,1)$.
The case  with $2$ worldsheet supersymmetries (known
as $\N=2$ string theory)
has been studied far less.  It was first
discussed in \cite{Ademollo:1976pp,Ademollo:1976an}.  Its target space meaning
was clarified in \cite{Ooguri:1991fp},
where it was shown that the closed string sector leads to self-dual gravity,
and in \cite{Ooguri:1991ie,Marcus:1992xt} where it was shown that
the heterotic or open string sector corresponds to self-dual
Yang-Mills,
in four dimensions with signature $++--$.  Moreover, it was
shown that for this string theory the infinite tower of string oscillations are absent from the
physical spectrum (they are BRST trivial).   This is what one
would ordinarily expect from a non-critical string theory, such
as $c\leq 1$ bosonic strings.  
Indeed, in a sense, for $\N=2$
strings the ``non-critical'' theories are equivalent to ``critical''
theories (the analog of the gap $1<c<25$ for bosonic string
is absent for $\N=2$ strings).

On the other hand, topological strings on Calabi-Yau 3-folds, or more 
precisely $\N=2$ topological strings, were introduced 
in \cite{Witten:1988xj}.    
The construction of the $\N=2$ topological string was
modeled after bosonic string theory, namely, a
twisted $\N=2$ worldsheet supersymmetry was used to construct a bosonic-string-like 
BRST complex.  This $\N=2$ topological string has two versions 
\cite{Vafa:1991uz,Witten:1991zz}, known as the A and B models, 
exchanged by mirror symmetry. 
It was shown in \cite{Witten:1992fb}
that the open topological strings in the A version
lead to ordinary Chern-Simons theory on Lagrangian
submanifolds of a Calabi-Yau space, while the B version leads to 
holomorphic Chern-Simons theory on the full Calabi-Yau.  The corresponding
closed string theory in the B model was studied in \cite{Bershadsky:1994cx},
leading to a gravity theory quantizing complex structure
of Calabi-Yau threefolds.  In any case, the target space description
always involves only finitely many fields.

Since the $\N=2$ string also involves only finitely many target space fields,
it was natural to expect \cite{Bershadsky:1994cx}
that the $\N=2$ string is also related to some kind of topological string; 
this was indeed shown to be the case in \cite{Berkovits:1995vy}.  The theory in question is the
$\N=4$ topological string theory, so named because it 
is connected to a twisted version of 
$\N=4$ worldsheet supersymmetry.  Moreover, this formulation leads
to vanishing theorems \cite{Berkovits:1995vy} for 
perturbative $\N=2$ string amplitudes for 
$n>3$ points on flat space at all loops.  It is also useful 
computationally, leading in certain cases
to computations of amplitudes at all loops \cite{Ooguri:1995cp}.

Thus both the $\N=2$ and the $\N=4$ topological strings
emerged as string theories for which string oscillations 
are absent from the physical
spectrum.  It was suggested in \cite{Bershadsky:1994cx} (pages 140-141) that these
two theories are indeed also related to each other.  
The motivation for this was that in the open ${\N }=2$ topological string one finds 
holomorphic Chern-Simons theory on a three complex dimensional
Calabi-Yau space (topological B model), while on the other
hand $\N=2$ strings, via the twistor transform, map
self-dual Yang-Mills geometry to
holomorphic bundles in twistor space \cite{Ward:1977ta}
which is also of complex dimension 3.  However, there was an obstacle
to realizing this idea: the twistor space is $\C\PP^3$ which
is {\it not} a Calabi-Yau threefold.

A few ideas emerged to resolve this puzzle \cite{Siegel:1992za,Siegel:1993wd,Chalmers:2000bh}:  
$\N=2$ strings naturally lead
to a theory in 2 complex dimensions \cite{Ooguri:1991fp}, with Lorentz
group $U(1,1)$.  Although this is not necessarily a problem,
it was suggested in \cite{Siegel:1992za} that one could try to
recover Lorentz invariance by ``integrating'' over the choices
of $U(1,1)\subset SO(2,2)$, which leads to bringing in the twistor 
sphere
as part of the physical space.  
However, it was found that this was not possible unless one also
added fermions and supersymmetrized the theory to $\N=4$
in the target space.  This echoes the fact that
$\C\PP^3$ is not a Calabi-Yau manifold but $\CPP$
is, as was recently exploited in \cite{Witten:2003nn}; 
thus there is a possibility to unify the two topological 
strings, as also discussed by \cite{berkovits-talk,Chalmers:2000bh}.
We believe that the results of \cite{Witten:2003nn} should 
be viewed as a step toward such a unification.  

Together with the recent observations in \cite{top-integ}
that $\N=2$ topological strings are equivalent to non-critical
bosonic strings and superstrings, we can view the equivalence of
$\N=2$ strings (which can also be viewed as ``non-critical'') with $\N=2$
topological strings as an extension of the principle
that all non-critical strings are unified by topological strings\footnote{
It is amusing to note that the dimension of the
critical topological string is correlated with the
upper limit of the central charge
of non-critical strings.  However, depending on how the twisting
is done there are two choices for the $\hat c$ of the corresponding $\N=2$ topological
theory \cite{Bershadsky:1993pe}.  
In particular, the $c=1$ non-critical
bosonic string gets mapped to $\hat c=3$ and $\hat c=-1$, which are precisely
the cases of most interest in topological strings (the latter being explored
here and in \cite{Witten:2003nn}).}.

\section{Brief review of A and B models}

In this section we provide a brief review of A and B model
strings on Calabi-Yau manifolds.  The reader
can consult \cite{MR2003030}
for a more thorough review.  To make the discussion
simple we first discuss the case which has been most studied, namely
the case where the Calabi-Yau is an ordinary (bosonic) manifold.

These topological models are defined by topological twisting of the $\N=2$ 
supersymmetry on the worldsheet of the 
supersymmetric sigma model with Calabi-Yau target space.  
In the closed string case, this twisting just amounts to shifting the spins of all 
worldsheet fields by $J \to J + \alpha/2$, where choosing $\alpha$ to be the 
$U(1)_R$ or $U(1)_A$ 
charge defines the A or B model respectively.  After this twist one of the two 
supercharges becomes scalar and can therefore be used as a BRST operator.
Correlation functions of BRST-invariant operators are then independent of worldsheet position, 
and depend on the background
in a very restricted way --- namely they depend only on the K\"ahler moduli (A model)
or complex moduli (B model).

Most importantly for our purposes, the fermionic 
BRST symmetry acting on the sigma model field space implies that correlation functions
are localized on BRST-invariant field configurations.  In the A model these are holomorphic
maps of the worldsheet into the target (worldsheet instantons), while in the B model they 
are simply the constant maps.  In this sense the B model reduces to classical geometry 
of points while the A model admits more complicated corrections from holomorphic curves.

\subsection{Open string case}

The A and B models also admit BRST-invariant boundary conditions for open strings.  
In the A model such a boundary condition is obtained by restricting the endpoints 
of the string to lie on a Lagrangian subspace $Y$ of the Calabi-Yau manifold $X$.  
In the B model the endpoints are coupled instead to a holomorphic bundle, supported 
on some holomorphic cycle inside the Calabi-Yau.  

The corresponding open string field theories were described 
in \cite{Witten:1992fb}.  Let us first discuss the A model.  In this case the BRST cohomology on
the worldsheet gets identified with the ordinary de Rham cohomology on $Y$.  Therefore the
target space 
fields can be packaged together as 
\begin{equation}
A = A_0 + A_1 + \cdots + A_m.
\end{equation} 
Here $A_k$ is a $k$-form on $Y$, and $m$ is the real dimension of $Y$.

The target space action consists of two parts.  One part is the 
Chern-Simons theory on $Y$, with action
\begin{equation}
S_{CS} = \frac{1}{g_s} \int_Y \Tr \left( A \wedge dA + \frac{2}{3} A \wedge A \wedge A \right).
\end{equation}
This part of the action can be considered as coming from disc diagrams in which the worldsheet has
degenerated to a Feynman diagram; these configurations lie at the boundary of field space and were
called ``virtual instantons'' in \cite{Witten:1992fb}.  In this Chern-Simons theory the one-form
part of $A$, $A_1$, is viewed as a $U(N)$ connection (if we have
$N$ branes), while the other $A_k$ can be viewed as the 
usual ghost fields which arise in covariant quantization of Chern-Simons 
\cite{Axelrod:1991vq,Axelrod:1994wr}.

The second contribution to the action comes from honest holomorphic maps 
$(\Sigma, \partial \Sigma) \to (X,Y)$.  Each such instanton modifies the 
action by adding a Wilson loop term, giving
\begin{equation}
S_{inst} =\sum_i C_i e^{-t_i}\ \Tr \ \mathrm{Pexp} \oint_{\gamma_i} A,
\end{equation} 
where $t_i$ is the area of the $i$-th
instanton and $\gamma_i$ is its boundary inside $Y$.  The coefficients
$C_i$ will depend on the details of the map.  The full target space
field theory can thus be viewed as the Chern-Simons theory coupled to a 
background source of Wilson lines, determined by the worldsheet
instantons.  Thus the full action is given by
\begin{equation}
S=S_{CS}+S_{inst}.
\end{equation}

The story is generally similar in the B model, 
with one important difference.  In this model, 
as we described above, the only
BRST invariant field configurations are the constant maps.  
Therefore there are no worldsheet
instanton corrections; the full open string theory 
for any Calabi-Yau target space 
is reduced to a holomorphic version of Chern-Simons, with action
\begin{equation}
S = \frac{1}{g_s} \int_X \Omega \wedge \Tr \left(A \wedge dA + 
\frac{2}{3} A \wedge A \wedge A \right).
\end{equation}
For the supermanifold case the same formulations
should work, although there are important technical
issues to be overcome, 
as has been recently discussed in the context of the 
B model in $\CPP$ \cite{Witten:2003nn}.

\section{Extra branes for A and B models?}
In this section we discuss the possible existence of extra branes
in the A and B models.  We will be mainly focusing on
the case of a bosonic Calabi-Yau 3-fold, but it should
be possible to generalize our discussion 
to supermanifolds such as $\CPP$.

Let us start with the B model.  This model has D-branes
of all even (Euclidean) dimensions.  In particular
we have D1-branes.  It has been known that D1-branes
play a privileged role among the branes of the B model 
\cite{Aganagic:2003db,top-integ}:  they are a source for the holomorphic
3-form.  In other words, if we have a 3-cycle $C_3$ surrounding
$N$ D1-branes, the period of the holomorphic 3-form $\Omega$
over this 3-cycle is deformed by the branes,
\begin{equation} \label{deform}
\Delta \int_{C_3} \Omega =N g_s.
\end{equation}
It is natural to introduce the 3-form
\begin{equation}
C=\frac{\Omega}{g_s}.
\end{equation}
Then \eqref{deform} can be written as
\begin{equation} \label{deform-2}
\Delta \int_{C_3} C = N.
\end{equation}
There is a better way to write this relation.  Let us introduce
a 2-form field $B$ under which the D1-brane is charged\footnote{
The need for such a field in the gravity sector of 
the twistorial Calabi-Yau was 
suggested in \cite{witten-priv}.}.
Then the above relation can be summarized by saying that the
action takes the form
\begin{equation} \label{action-b-model}
S=S_{KS}+\int C\wedge dB,
\end{equation}
where $S_{KS}$ is the Kodaira-Spencer action 
\cite{Bershadsky:1994cx}.  One can easily
check that the addition of this term leads,
upon variation of the Kodaira-Spencer
field, to \eqref{deform-2}.

So far we have been discussing the B model.
In the A model context the same kind of relation 
holds (and has been related to the large $N$ duality
between Chern-Simons and closed topological strings 
\cite{Gopakumar:1998vy}):  a D2-brane wrapped over a Lagrangian
3-cycle $Y$ changes the integral of the K\"ahler form $k$
over a 2-cycle surrounding $Y$.  In this case
we introduce a gauge potential $\tilde{C}$ which couples to the D2 brane,
and a normalized K\"ahler form
${\tilde B}=k/g_s$, and then the action includes
\begin{equation}
S=...+\int {\tilde B}\wedge d{\tilde C}.
\end{equation}
Note that this is exactly the same as \eqref{action-b-model}
with the identifications
\begin{eqnarray}
B\leftrightarrow {\tilde B}, \\
C\leftrightarrow {\tilde C}.
\end{eqnarray}
We will later propose that this fact is not accidental but
rather a consequence of S-duality, in
the specific example of $\CPP$.

It is natural to ask whether topological D-branes can
end on other branes.  Given that this does happen
in the superstring, one might expect that
the same thing also happens in the topological string
context.  In particular, the computation of F-terms
for superstring compactifications including branes ending
on branes should be captured by some topological theory.
For D1-branes, the only way
this can happen (consistent with holomorphy,
which is a consequence of the BRST symmetry of the B model)
is for the branes to end on Lagrangian submanifolds.  
What is the nature of these Lagrangian submanifolds?  They
certainly cannot be D-branes, since this would violate the BRST symmetry.
The Lagrangian submanifold must be supporting some new kind of brane 
of the B model; let us call this hypothetical brane an ``NS2-brane.''
If such a brane really exists in the B model,
then it should naturally couple to the gravity sector.
Since the gravity sector of the B model is rather poor in
fields, the only natural guess for the coupling is
\begin{equation}
c \int_Y \Omega.
\end{equation}
Later we will identify the constant $c$ with $1/g_s$.

Now can a D1-brane end on this NS-brane?
As is familiar from superstring theory, charge conservation
requires that if branes $Z$ and $Z'$ can end on
brane $Y$, the action must contain a Chern-Simons term
of the form
\begin{equation}
S_{CS}=\int A \wedge F \wedge F'.
\end{equation}
Here $F$ and $F'$ denote the field
strengths coupled to $Z$ and $Z'$, while $A$ is the corresponding gauge
potential for the field coupled to $Y$.
In the case at hand $Y$ is our Lagrangian NS-brane, $A$ is the 3-form $C$, 
and $F=dB$ is the 3-form field strength coupled to the D1.  Hence $F'$
should be a closed 0-form, i.e. a constant function.
So if D1-branes can end on NS2-branes, we would expect a term in the
action of the form $\int C\wedge dB$; but as we have argued
above, there {\it is} just such a term in the action. 
This is consistent with the assumption that these
Lagrangian NS2-branes exist and that D1-branes 
can end on them.

\section{Tadpoles on the twistorial Calabi-Yau and S-duality}

In this section we discuss possible gravitational
tadpoles for the B model
on $\CPP$ with space-filling D5-branes.  We argue how
such tadpoles can arise and propose possible cancellation
mechanisms.  We then discuss the relation between this picture
and the twistor transform of \cite{Witten:2003nn}.
Finally we
introduce the notion of S-duality for the topological string,
and suggest that this B model could in fact be 
S-dual to an A model on $\CPP$.

So, consider a B model with $N$ D5-branes wrapped over $\CPP$.  
In analogy with type I superstrings in 10 dimensions, we
know that there could very well be gravitational tadpoles.
A comprehensive study of such tadpoles would be important
for this theory.  Here we describe some possible tadpoles that
can exist.  One possibility would be a ``bulk'' tadpole ---
a gravitational mode which couples to the total brane charge,
much as in the type I superstring.  Another possibility
would be an ``induced'' tadpole which gives rise to a lower-dimensional
brane.

Let us first discuss the possibility of a bulk tadpole. 
In the case of bosonic
Calabi-Yau 3-folds, the only canonical choice would be
a brane generated
term of the form
\begin{equation}
\int \Omega \wedge {\overline \Omega}.
\end{equation}  
Even though
$\overline \Omega$ is not a field of the B model and
depends on anti-holomorphic moduli, one could
nevertheless countenance the existence of 
such a term because of the holomorphic anomaly 
\cite{Bershadsky:1994cx}.  
If such a term is indeed generated, the only known mechanism
in the superstring context
to cancel it is the introduction of orientifold planes.
Indeed, it is known that orientifold planes introduce
brane charge also in topological string theories \cite{Sinha:2000ap}.
In particular, the orientifold action which fixes the
full Calabi-Yau gives a brane charge of $\pm 8=\pm 2^3$ which leads
to the introduction of $8$ branes and an $SO(8)$ gauge group in the bulk.  
It is conceivable that a similar mechanism is at work
for Calabi-Yaus which are supermanifolds.  If so, this
would give a fixed rank gauge group.  It would be somewhat surprising,
however, since $\N=4$ super Yang-Mills seems to be well-behaved and
non-anomalous independent of the chosen gauge group.\footnote{Also, 
for the branes of \cite{Witten:2003nn} the orientifold
action presumably would fix just the brane, which is actually not
quite space-filling; it has only
4 real fermionic directions.  In such a case the analog
of $\int \Omega \wedge {\overline \Omega}$ is not clear.
}

For now let us assume either that there are no
``bulk'' tadpoles, or at least that defining a 
decoupled gauge theory sector only requires
cancellations of lower dimensional``induced'' tadpoles.
Now we consider these.
As we have reviewed, on the worldvolume of D5 branes we have
the action
\begin{equation}
\frac{1}{g_s}\int_{\CPP} \Omega \wedge CS(A)
\end{equation}
One can also ask whether these branes
induce any gravitational backreaction, which would be present even if $A=0$.  
In particular,
since the topological string at the disc level
computes superpotential terms
for superstring compactifications on the Calabi-Yau 
\cite{Bershadsky:1994cx}, and since
we expect that superpotential to vanish when the spin connection $\omega$
is identified with the gauge connection $A$ 
\cite{Witten:1986bz}, one would expect
that the full disc contribution is
\begin{equation} \label{full-disc}
\frac{1}{g_s} \int_{\CPP} \Omega \wedge [CS(A)-CS(\omega)],
\end{equation}
where $\omega$ denotes the $(0,1)$ part of the spin connection.
The existence of
such additional gravitational Chern-Simons terms is also natural
from the perspective of ordinary Chern-Simons theory
\cite{Witten:1989hf}; there such a gravitational term is
needed to ensure topological invariance is not spoiled quantum mechanically.  
However,
in that case the gravitational Chern-Simons arises with a coefficient 
given by the central charge of the WZW model, which in the $U(N)_k$ case 
would have been $kN^2/(N+k)$.
In view of this it would be crucial to check in the string theory
under consideration what is the precise coefficient of the above term.
Let us assume that this term is indeed generated at the level of the 
disc amplitude as in \eqref{full-disc}.

To write \eqref{full-disc} in a more convenient form, note that $CS(\omega)$
should be related to a Poincare dual
$(3|4)$ cycle.  While a full theory of ``superhomology'' is apparently
lacking, there is one natural candidate for this cycle.  Namely,
choosing an antiholomorphic involution on $\CPP$, the fixed locus gives
an $\RPP$, which on general grounds should be a special Lagrangian cycle
and in particular should be nontrivial in homology.  Let us assume that
this candidate cycle is indeed dual to $CS(\omega)$ (with coefficient $1$, say,
though the exact coefficient is inessential.)
Then we would have\footnote{We are glossing 
over some formidable mathematical subtleties here.  As usually defined, the 
holomorphic Chern-Simons action is
not gauge invariant and really makes sense only modulo the periods of $\Omega$ ---
see e.g. \cite{Thomas:1998uj}.  Nevertheless the question of whether or not there is a tadpole should 
have a well-defined answer.}
\begin{equation} \label{tadpole}
\frac{-1}{g_s}\int_{\CPP}\Omega \wedge CS(\omega)=\frac{-1}{g_s}
\int_{\RPP} \Omega.
\end{equation}
Recall that $\Omega$ is the Kodaira-Spencer
field which describes the gravitational sector of the B model 
\cite{Bershadsky:1994cx}.  Thus \eqref{tadpole} can be interpreted
as a gravitational tadpole created by a single D5-brane.  Letting $C = 
\Omega / g_s$, for $N$ D5-branes we have
\begin{equation} \label{tadpole-multi}
-N\int_{\RPP} C.
\end{equation}
So the only way this B model with D5-branes can be consistent is if it has
an additional term in the action involving $C$.
But as we have argued in the previous section, there is a candidate:
namely, we proposed that the B model could include an NS2-brane
coupled to $C$.  Moreover,
we have argued that, if NS2-branes do exist, 
D1-branes can end on them.  We are
thus led to introduce $N$ of these NS2-branes
wrapping the Lagrangian cycle $\RPP$, which would
cancel the tadpole \eqref{tadpole-multi}.

\subsection{Twistor transform of $\N=4$ super Yang-Mills revisited}

After our discussion above, we can now revisit the D1-brane instantons studied
in \cite{Witten:2003nn}.  Since
D1-branes can end on the NS2-branes, we obtain 
a natural real cycle inside each moduli space of D1-branes.
The geometrical picture is that the closed
Riemann surfaces considered in \cite{Witten:2003nn} should be
considered as doubled versions
of the open D1-branes ending on $\RPP$.  The key point is that not every 
closed Riemann surface
is obtained in this way.  To characterize the ones which {\it are} obtained by doubling,
note that the antiholomorphic involution which acts on $\CPP$ with fixed locus
$\RPP$ also acts naturally on the instanton moduli space.  The instantons obtained
by doubling
are exactly the fixed points of this involution; these form a real cycle in the
moduli space as needed.

In particular, the contributions
considered in \cite{Witten:2003nn} with genus $g$ get 
mapped in our picture to D1-instantons with no handles and 
with $b=g+1$ boundaries ending on $\RPP$.  This is exactly the same
topology as the corresponding $\N=4$ `t Hooft diagram
the instanton is computing!  This is also consistent with 
the fact that each boundary of the instanton contributes 
a factor of $N$, coming from the degeneracy of the 
Lagrangian branes on which it ends,
as would be required for the corresponding gauge theory
diagram.  Furthermore, this counting suggests the existence
of a constant background field $g_{D1}$ coupled to the curvature of the
$D1$ brane, which we identify with the square of the Yang-Mills coupling
constant:
\begin{equation}
g_{YM}^2=g_{D1}.
\end{equation}

It is natural to ask whether $g_{D1}$ is
the same as the B-model topological string coupling $g_s$.  This would
be reasonable in view of the fact that the tree level amplitude
includes a $1/g_s$, which in the Yang-Mills theory should have
corresponded to a $1/g_{YM}^2=1/g_{D1}$.

At first it seems however
that this agreement is spurious, because a D1-brane
instanton of degree $d$ should get an additional weight
$e^{-Ad/g_s}$, where $A$ is 
the area of the degree 1 curve.  This
extra weighting can actually be absorbed using 
the condition given in \cite{Witten:2003nn} for instanton
contributions to planar Yang-Mills amplitudes:
\begin{equation} \label{degree-constraint}
d = q + l - 1.
\end{equation}
Here $q$ is the number of ingoing negative helicity gluons
in the Yang-Mills scattering process, and $l$ 
is the number of loops in Yang-Mills which for
planar diagrams is the same as the $l=b-1$ where $b$
is the number of boundaries.
In principle we can consider an instanton with an arbitrary
number of handles $h$, and
$b$ boundaries lying on $\RPP$.  We propose that the generalization of
this formula to arbitrary $h$ is given by
\begin{equation}
d=q+2h+b-2
\end{equation}
Now we can see that we can absorb
the factor $e^{-Ad/g_s}$ in two rescalings:  we rescale
\begin{equation}
g_{D1} \to e^{A/g_s} g_{D1},
\end{equation}
and also rescale the vertex operators $V_-$ for all negative helicity gluons by 
\begin{equation}
V_- \to e^{A/g_s} V_-.
\end{equation}
This triviality of the K\"ahler class is reminiscent of \cite{Sethi:1994ch}, where it was shown that the K\"ahler form
is indeed decoupled from the BRST invariant observables in 
certain A models on supermanifolds.  

Thus it is plausible that we can indeed set $g_s=g_{D1}$.  At any rate, for the rest
of the paper we need not assume this.

Note that there could be a new puzzle associated with the
$N$ additional NS2-branes wrapping $\RPP$.  If these branes
carry additional dynamical fields, it will 
ruin the perfect match with the field multiplet of $\N=4$ super Yang-Mills,
which was obtained in \cite{Witten:2003nn} just from the fields on the
D5-branes.
Clearly, a $U(N)$ gauge field lives on the NS2-branes (due to the fact
that D1-branes can end on them).  So we expect a 
$U(N)$ theory with no dynamical fields.  
The natural candidate is the Chern-Simons
theory living on the branes; on $\RPP$ the space of classical
solutions of this theory is purely discrete.

\subsection{S-duality and the A model}

The D1 brane instantons we have been discussing are quite similar
to the fundamental strings of the A model.  In particular,
the fact that they end on Lagrangian branes is reminiscent of 
fundamental strings ending on
the D-branes of the A model.  Furthermore, we have argued above 
that consistency requires these branes to support Chern-Simons 
gauge theory.  It is thus natural to suggest that there is some 
dual description in which the A model picture is the right one.
In this picture the ``NS2-branes'' get mapped to D2-branes
and D1-branes get mapped to fundamental strings.

Independently, it is natural to ask how the Montonen-Olive
duality $g_{YM} \leftrightarrow 1/g_{YM}$ acts on this picture.
This question exists even if we have no tadpoles to cancel
and the D1 instantons end up being closed Riemann surfaces.
A priori there could be two natural answers.  One possibility 
is that the picture is mapped back to itself.  This we find unlikely, given
the fact that the D1 string coupling constant is identified
with $g_{YM}^2$, and as in type IIB duality we would have
expected the duality to map the D1 string to the 
fundamental string.

We thus conjecture that the S-dual description of this model
is in terms of A model topological strings, with $N$ D-branes
wrapping $\RPP$.  However, this A model is novel in that it
should also have $N$ additional ``NS5-branes.''
A correct quantization of the 
full system including these extra NS5-branes could then avoid the 
A model no-go theorem \cite{Witten:2003nn}.

Actually, the fact that the A model can be related to D-instantons
has already been encountered \cite{Gopakumar:1998ii,Gopakumar:1998jq}.
In particular, the worldsheet instantons of the A model
capture degeneracies of wrapped D2 branes, which can
be viewed as Euclidean D1 brane instantons.  This has also been
extended to a relation between the degeneracy
of D2 branes ending on branes and A model
worldsheets ending on Lagrangian D-branes \cite{Ooguri:1999bv}.
If we want to interpret these computations completely within the
topological string we would need an S-duality like the one we
are proposing.

It would be natural to try to define the amplitudes
involving instantons ending on branes directly from the A model
perspective. By now
there is a huge machinery for such computations
for bosonic toric Calabi-Yau threefolds.  The Calabi-Yau space 
$\CPP$ is also toric, albeit super, and one would expect
that suitable localization techniques could also work
here.

\section{Mirror symmetry}

So far we have discussed a possible stringy twistorial formulation of 
$\N=4$ supersymmetric $U(N)$ Yang-Mills coupled to $\N=4$
conformal supergravity; in fact we proposed that there are two such formulations,
both as A and B model topological string on $\CPP$, in each case with $N$ space-filling
branes and $N$ branes
wrapping $\RPP$.  We have seen
that instantons do contribute to various
amplitudes in the twistor space.  In the A-model case, the
instantons are worldsheet instantons.
As is well known, the main
point of mirror symmetry is to compute precisely
such worldsheet instanton contributions.  Namely, in the mirror 
(not S-dual) B model
description the instantons disappear; the field theory in
the B model captures the worldsheet instantons of the A model.
This strategy has been successfully applied to Calabi-Yau
threefolds in the presence of Lagrangian branes, starting
with \cite{Aganagic:2000gs} and culminating in
a recent work \cite{top-integ}.  Therefore, we would expect this 
strategy to work here as well.\footnote{For a
recent discussion of mirror symmetry for $\N=2$ strings
see \cite{Cheung:2002yw}.}

How does mirror symmetry work for Calabi-Yau supermanifolds?
In fact Calabi-Yau supermanifolds were first studied precisely
for a better understanding of mirror symmetry \cite{Sethi:1994ch}!
Namely, the puzzle that rigid Calabi-Yaus cannot have ordinary
Calabi-Yau mirrors (as the mirror would not have any K\"ahler classes)
was resolved in \cite{Sethi:1994ch} by extending
the space of bosonic Calabi-Yaus to include Calabi-Yau supermanifolds.
In particular, examples were found where the mirror of a rigid Calabi-Yau
is a Calabi-Yau supermanifold, realized as a complete intersection in a
product of projective supermanifolds $\prod_i {\bf CP}^{n_i|m_i}$.  In 
such cases the superdimension of the manifold and the mirror agree
(which is required because the superdimension determines the $\hat c$ in the $U(1)$ 
current algebra), but the individual bosonic and fermionic dimenions
need not be equal.  In particular it was found in \cite{Sethi:1994ch}
that a Calabi-Yau of dimension $(n|0)$ can be mirror to
a Calabi-Yau with dimension $(n+d|d)$. 

The techniques to derive mirror geometry from the linear
sigma model description have been found in \cite{Hori:2000kt}
in the context of bosonic Calabi-Yaus.  To treat $\CPP$ one would need to extend
these techniques to the case of Calabi-Yau supermanifolds.
This should be possible, given that $\CPP$
does admit a simple linear sigma model realization.
Here we would like to make a conjecture about what the mirror is.
Whatever it is, it will carry the holomorphic version of Chern-Simons.  
Moreover, via a twistor transform, it should realize Yang-Mills perturbation theory 
directly from holomorphic Chern-Simons, since that theory has no corrections from 
worldsheet instantons.  Therefore the full classical
Yang-Mills should be realized completely in a classical way after the 
twistor transform.  There is only one such formulation presently
known:  classical solutions of the full Yang-Mills can be identified with
holomorphic data on the quadric in $\C\PP^{3|3}\times\C\PP^{3|3}$ 
\cite{Witten:1978xx}.  We thus conjecture
that the mirror of $\CPP$ is a quadric in ${\bf CP}^{3|3}\times
{\bf CP}^{3|3}$.  First of all, note that this quadric is a Calabi-Yau 
supermanifold.  This is already a non-trivial
test. In fact, it was already conjectured in \cite{Witten:2003nn}
that the B model on this quadric could reproduce $\N=4$ super Yang-Mills. Furthermore, both manifolds have the same superdimension, $-1$, as required by mirror symmetry. 

We are currently investigating the validity of this mirror conjecture.
If the conjecture is proven it would be an important step toward explaining,
more or less from first principles, the fact that the twistor
transformed amplitudes lie on holomorphic curves; it is because
the A model on $\CPP$ is the mirror of a B model on twistor space whose
perturbation theory is the same as that of ordinary 
$\N=4$ super Yang-Mills.

\section{Conclusion}

In this paper we have presented some ideas which 
could lead to a different understanding
of the twistor transform of $\N=4$ super Yang-Mills presented in \cite{Witten:2003nn}.
Many of our ideas are clearly rather speculative; they should be
viewed as broadening the possibilities for a consistent stringy realization
of the beautiful ideas of the twistor transform \cite{Penrose:1976js,Ward:1977ta}.

Quite independent of the application to
$\N=4$ super Yang-Mills, it would be very desirable to 
understand whether the new NS-branes we proposed in the A and B models indeed exist,
and whether there is indeed an S-duality exchanging D-branes with NS-branes.  These
questions would make sense even if it turns out that there are no gravitational tadpoles. 
In particular, our S-duality conjecture relating the A and B model on the same Calabi-Yau
and the mirror conjecture we made in the previous section
could have a broader range of validity.

More optimistically, if our picture turns out to be correct,
it would be important to understand carefully how the physical states of $\N=4$ super 
Yang-Mills can be encoded in the A model with extra NS5-branes --- in other words, what 
is the precise origin in this picture of the Dolbeault cohomology which appears in the B model?  
In the A model the computation of $\N=4$ super Yang-Mills amplitudes should amount to some version 
of worldsheet perturbation theory, and the issue is to determine the proper vertex operators to 
insert.  It should then be possible to calculate $\N=4$ super Yang-Mills 
amplitudes in the A model; in particular one should be able to 
obtain the maximally helicity-violating amplitudes, 
as was done in \cite{Witten:2003nn} for the B model.
One
would imagine that the effects of the NS5-branes can be taken into account
by a suitable definition of the measure on
 the moduli space of open A model instantons.
This would be a crucial test of our proposed A model picture.

\section*{Acknowledgements}

We thank Nathan Berkovits,
Sergei Gukov, Sheldon Katz, Lubo\v{s} Motl, Hirosi Ooguri, Warren Siegel and Edward Witten for 
several helpful conversations.
This research is supported in part by NSF grants PHY-0244821 and DMS-0244464.

\renewcommand{\baselinestretch}{1}
\small\normalsize

%\bibliography{physics}

\begin{thebibliography}{10}

\bibitem{Witten:2003nn}
E.~Witten, ``{P}erturbative gauge theory as a string theory in twistor space,''
\href{http://www.arXiv.org/abs/hep-th/0312171}{{\tt hep-th/0312171}}.
%%CITATION = HEP-TH 0312171;%%.

\bibitem{Roiban:2004vt}
R.~Roiban, M.~Spradlin, and A.~Volovich, ``{A} googly amplitude from the
  {B}-model in twistor space,''
\href{http://www.arXiv.org/abs/hep-th/0402016}{{\tt hep-th/0402016}}.
%%CITATION = HEP-TH 0402016;%%.

\bibitem{Berkovits:2004hg}
N.~Berkovits, ``{A}n alternative string theory in twistor space for {$\N=4$}
  super-{Y}ang-{M}ills,''
\href{http://www.arXiv.org/abs/hep-th/0402045}{{\tt hep-th/0402045}}.
%%CITATION = HEP-TH 0402045;%%.

\bibitem{Roiban:2004ka}
R.~Roiban and A.~Volovich, ``{A}ll googly amplitudes from the {B}-model in
  twistor space,''
\href{http://www.arXiv.org/abs/hep-th/0402121}{{\tt hep-th/0402121}}.
%%CITATION = HEP-TH 0402121;%%.

\bibitem{Parke:1985pn}
S.~J. Parke and T.~R. Taylor, ``{P}erturbative {QCD} utilizing extended
  supersymmetry,'' {\em Phys. Lett.} {\bf B157} (1985)
81.
%%CITATION = PHLTA,B157,81;%%.

\bibitem{Parke:1986gb}
S.~J. Parke and T.~R. Taylor, ``{A}n amplitude for {$n$} gluon scattering,''
  {\em Phys. Rev. Lett.} {\bf 56} (1986)
2459.
%%CITATION = PRLTA,56,2459;%%.

\bibitem{Bern:1993mq}
Z.~Bern, L.~J. Dixon, and D.~A. Kosower, ``One loop corrections to five gluon
  amplitudes,'' {\em Phys. Rev. Lett.} {\bf 70} (1993) 2677--2680,
\href{http://www.arXiv.org/abs/hep-ph/9302280}{{\tt hep-ph/9302280}}.
%%CITATION = HEP-PH 9302280;%%.

\bibitem{Bern:2002tk}
Z.~Bern, A.~De~Freitas, and L.~J. Dixon, ``{T}wo-loop helicity amplitudes for
  gluon gluon scattering in {QCD} and supersymmetric {Y}ang-{M}ills theory,''
  {\em JHEP} {\bf 03} (2002) 018,
\href{http://www.arXiv.org/abs/hep-ph/0201161}{{\tt hep-ph/0201161}}.
%%CITATION = HEP-PH 0201161;%%.

\bibitem{Bern:2003ck}
Z.~Bern, A.~De~Freitas, and L.~J. Dixon, ``Two-loop helicity amplitudes for
  quark gluon scattering in {QCD} and gluino gluon scattering in supersymmetric
  {Y}ang-{M}ills theory,'' {\em JHEP} {\bf 06} (2003) 028,
\href{http://www.arXiv.org/abs/hep-ph/0304168}{{\tt hep-ph/0304168}}.
%%CITATION = HEP-PH 0304168;%%.

\bibitem{Nair:1988bq}
V.~P. Nair, ``A current algebra for some gauge theory amplitudes,'' {\em Phys.
  Lett.} {\bf B214} (1988)
215.
%%CITATION = PHLTA,B214,215;%%.

\bibitem{'tHooft:1974jz}
G.~'t~Hooft, ``A planar diagram theory for strong interactions,'' {\em Nucl.
  Phys.} {\bf B72} (1974)
461.
%%CITATION = NUPHA,B72,461;%%.

\bibitem{Ooguri:1991fp}
H.~Ooguri and C.~Vafa, ``Geometry of {$\N=2$} strings,'' {\em Nucl. Phys.} {\bf
  B361} (1991)
469--518.
%%CITATION = NUPHA,B361,469;%%.

\bibitem{Witten:1978xx}
E.~Witten, ``{A}n interpretation of classical {Y}ang-{M}ills theory,'' {\em
  Phys. Lett.} {\bf B77} (1978)
394.
%%CITATION = PHLTA,B77,394;%%.

\bibitem{Ketov:1993ix}
S.~J. Gates, S.~V. Ketov, and H.~Nishino, ``{S}elfdual supersymmetry and
  supergravity in {A}tiyah-{W}ard space-time,'' {\em Nucl. Phys.} {\bf B393}
  (1993) 149--210,
\href{http://www.arXiv.org/abs/hep-th/9207042}{{\tt hep-th/9207042}}.
%%CITATION = HEP-TH 9207042;%%.

\bibitem{Okounkov:2003sp}
A.~Okounkov, N.~Reshetikhin, and C.~Vafa, ``{Q}uantum {C}alabi-{Y}au and
  classical crystals,''
\href{http://www.arXiv.org/abs/hep-th/0309208}{{\tt hep-th/0309208}}.
%%CITATION = HEP-TH 0309208;%%.

\bibitem{Iqbal:2003ds}
A.~Iqbal, N.~Nekrasov, A.~Okounkov, and C.~Vafa, ``Quantum foam and topological
  strings,''
\href{http://www.arXiv.org/abs/hep-th/0312022}{{\tt hep-th/0312022}}.
%%CITATION = HEP-TH 0312022;%%.

\bibitem{Ademollo:1976pp}
M.~Ademollo et al, ``Dual string with {$U(1)$} color symmetry,'' {\em Nucl.
  Phys.} {\bf B111} (1976)
77--110.
%%CITATION = NUPHA,B111,77;%%.

\bibitem{Ademollo:1976an}
M.~Ademollo et al, ``Supersymmetric strings and color confinement,'' {\em Phys.
  Lett.} {\bf B62} (1976)
105.
%%CITATION = PHLTA,B62,105;%%.

\bibitem{Ooguri:1991ie}
H.~Ooguri and C.~Vafa, ``{$\N=2$} heterotic strings,'' {\em Nucl. Phys.} {\bf
  B367} (1991)
83--104.
%%CITATION = NUPHA,B367,83;%%.

\bibitem{Marcus:1992xt}
N.~Marcus, ``The {$\N=2$} open string,'' {\em Nucl. Phys.} {\bf B387} (1992)
  263--279,
\href{http://www.arXiv.org/abs/hep-th/9207024}{{\tt hep-th/9207024}}.
%%CITATION = HEP-TH 9207024;%%.

\bibitem{Witten:1988xj}
E.~Witten, ``Topological sigma models,'' {\em Commun. Math. Phys.} {\bf 118}
  (1988)
411.
%%CITATION = CMPHA,118,411;%%.

\bibitem{Vafa:1991uz}
C.~Vafa, ``Topological mirrors and quantum rings,''
  \href{http://www.arXiv.org/abs/hep-th/9111017}{{\tt hep-th/9111017}}.
In Yau, S.T. (ed.): {\textit {Mirror symmetry I}}, 97-120.
%%CITATION = HEP-TH 9111017;%%.

\bibitem{Witten:1991zz}
E.~Witten, ``Mirror manifolds and topological field theory,''
  \href{http://www.arXiv.org/abs/hep-th/9112056}{{\tt hep-th/9112056}}.
In Yau, S.T. (ed.): {\textit{Mirror symmetry I}}, 121-160.
%%CITATION = HEP-TH 9112056;%%.

\bibitem{Witten:1992fb}
E.~Witten, ``{Chern-Simons} gauge theory as a string theory,'' {\em Prog.
  Math.} {\bf 133} (1995) 637--678,
\href{http://www.arXiv.org/abs/hep-th/9207094}{{\tt hep-th/9207094}}.
%%CITATION = HEP-TH 9207094;%%.

\bibitem{Bershadsky:1994cx}
M.~Bershadsky, S.~Cecotti, H.~Ooguri, and C.~Vafa, ``Kodaira-{S}pencer theory
  of gravity and exact results for quantum string amplitudes,'' {\em Commun.
  Math. Phys.} {\bf 165} (1994) 311--428,
\href{http://www.arXiv.org/abs/hep-th/9309140}{{\tt hep-th/9309140}}.
%%CITATION = CMPHA,165,311;%%.

\bibitem{Berkovits:1995vy}
N.~Berkovits and C.~Vafa, ``{$\N=4$} topological strings,'' {\em Nucl. Phys.}
  {\bf B433} (1995) 123--180,
\href{http://www.arXiv.org/abs/hep-th/9407190}{{\tt hep-th/9407190}}.
%%CITATION = HEP-TH 9407190;%%.

\bibitem{Ooguri:1995cp}
H.~Ooguri and C.~Vafa, ``All loop {$\N=2$} string amplitudes,'' {\em Nucl.
  Phys.} {\bf B451} (1995) 121--161,
\href{http://arXiv.org/abs/hep-th/9505183}{{\tt hep-th/9505183}}.
%%CITATION = HEP-TH 9505183;%%.

\bibitem{Ward:1977ta}
R.~S. Ward, ``{O}n selfdual gauge fields,'' {\em Phys. Lett.} {\bf A61} (1977)
81--82.
%%CITATION = PHLTA,A61,81;%%.

\bibitem{Siegel:1992za}
W.~Siegel, ``The {$\N=2$} (4) string is selfdual {$\N=4$} {Y}ang-{M}ills,''
\href{http://www.arXiv.org/abs/hep-th/9205075}{{\tt hep-th/9205075}}.
%%CITATION = HEP-TH 9205075;%%.

\bibitem{Siegel:1993wd}
W.~Siegel, ``Selfdual {$\N=8$} supergravity as closed {$\N=2$} ({$\N=4$})
  strings,'' {\em Phys. Rev.} {\bf D47} (1993) 2504--2511,
\href{http://www.arXiv.org/abs/hep-th/9207043}{{\tt hep-th/9207043}}.
%%CITATION = HEP-TH 9207043;%%.

\bibitem{Chalmers:2000bh}
G.~Chalmers and W.~Siegel, ``Global conformal anomaly in {$\N = 2$} string,''
  {\em Phys. Rev.} {\bf D64} (2001) 026001,
\href{http://www.arXiv.org/abs/hep-th/0010238}{{\tt hep-th/0010238}}.
%%CITATION = HEP-TH 0010238;%%.

\bibitem{berkovits-talk}
N.~Berkovits.
\newblock {T}alk at STRINGS 2003, {K}yoto.

\bibitem{top-integ}
M.~Aganagic, R.~Dijkgraaf, A.~Klemm, M.~Marino, and C.~Vafa, ``{T}opological
  strings and integrable hierarchies,''
  \href{http://www.arXiv.org/abs/hep-th/0312085}{{\tt hep-th/0312085}}.

\bibitem{Bershadsky:1993pe}
M.~Bershadsky, W.~Lerche, D.~Nemeschansky, and N.~P. Warner, ``Extended
  {$\N=2$} superconformal structure of gravity and {W} gravity coupled to
  matter,'' {\em Nucl. Phys.} {\bf B401} (1993) 304--347,
\href{http://www.arXiv.org/abs/hep-th/9211040}{{\tt hep-th/9211040}}.
%%CITATION = HEP-TH 9211040;%%.

\bibitem{MR2003030}
K.~Hori, S.~Katz, A.~Klemm, R.~Pandharipande, R.~P. Thomas, C.~Vafa, R.~Vakil,
  and E.~Zaslow, {\em Mirror symmetry}, vol.~1 of {\em Clay Mathematics
  Monographs}.
\newblock American Mathematical Society, Providence, RI, 2003.

\bibitem{Axelrod:1991vq}
S.~Axelrod and I.~M. Singer, ``{C}hern-{S}imons perturbation theory,''
  \href{http://www.arXiv.org/abs/hep-th/9110056}{{\tt hep-th/9110056}}.
In {\textit{Proceedings, Differential geometric methods in theoretical
  physics}}, vol. 1, 3-45.
%%CITATION = HEP-TH 9110056;%%.

\bibitem{Axelrod:1994wr}
S.~Axelrod and I.~M. Singer, ``{C}hern-{S}imons perturbation theory. 2,'' {\em
  J. Diff. Geom.} {\bf 39} (1994) 173--213,
\href{http://www.arXiv.org/abs/hep-th/9304087}{{\tt hep-th/9304087}}.
%%CITATION = HEP-TH 9304087;%%.

\bibitem{Aganagic:2003db}
M.~Aganagic, A.~Klemm, M.~Marino, and C.~Vafa, ``The topological vertex,''
\href{http://www.arXiv.org/abs/hep-th/0305132}{{\tt hep-th/0305132}}.
%%CITATION = HEP-TH 0305132;%%.

\bibitem{witten-priv}
E.~Witten.
\newblock Private communication.

\bibitem{Gopakumar:1998vy}
R.~Gopakumar and C.~Vafa, ``Topological gravity as large {$N$} topological
  gauge theory,'' {\em Adv. Theor. Math. Phys.} {\bf 2} (1998) 413--442,
\href{http://arXiv.org/abs/hep-th/9802016}{{\tt hep-th/9802016}}.
%%CITATION = HEP-TH 9802016;%%.

\bibitem{Sinha:2000ap}
S.~Sinha and C.~Vafa, ``{SO} and {S}p {C}hern-{S}imons at large {N},''
\href{http://www.arXiv.org/abs/hep-th/0012136}{{\tt hep-th/0012136}}.
%%CITATION = HEP-TH 0012136;%%.

\bibitem{Witten:1986bz}
E.~Witten, ``New issues in manifolds of ${S}{U}(3)$ holonomy,'' {\em Nucl.
  Phys.} {\bf B268} (1986)
79.
%%CITATION = NUPHA,B268,79;%%.

\bibitem{Witten:1989hf}
E.~Witten, ``Quantum field theory and the {J}ones polynomial,'' {\em Commun.
  Math. Phys.} {\bf 121} (1989)
351.
%%CITATION = CMPHA,121,351;%%.

\bibitem{Thomas:1998uj}
R.~P. Thomas, ``{A} holomorphic {C}asson invariant for {C}alabi-{Y}au 3-folds,
  and bundles on {K}3 fibrations,''
\href{http://www.arXiv.org/abs/math.ag/9806111}{{\tt math.ag/9806111}}.
%%CITATION = MATH.AG 9806111;%%.

\bibitem{Sethi:1994ch}
S.~Sethi, ``Supermanifolds, rigid manifolds and mirror symmetry,'' {\em Nucl.
  Phys.} {\bf B430} (1994) 31--50,
\href{http://www.arXiv.org/abs/hep-th/9404186}{{\tt hep-th/9404186}}.
%%CITATION = HEP-TH 9404186;%%.

\bibitem{Gopakumar:1998ii}
R.~Gopakumar and C.~Vafa, ``M-theory and topological strings. {I},''
\href{http://www.arXiv.org/abs/hep-th/9809187}{{\tt hep-th/9809187}}.
%%CITATION = HEP-TH 9809187;%%.

\bibitem{Gopakumar:1998jq}
R.~Gopakumar and C.~Vafa, ``M-theory and topological strings. {II},''
\href{http://www.arXiv.org/abs/hep-th/9812127}{{\tt hep-th/9812127}}.
%%CITATION = HEP-TH 9812127;%%.

\bibitem{Ooguri:1999bv}
H.~Ooguri and C.~Vafa, ``Knot invariants and topological strings,'' {\em Nucl.
  Phys.} {\bf B577} (2000) 419--438,
\href{http://arXiv.org/abs/hep-th/9912123}{{\tt hep-th/9912123}}.
%%CITATION = HEP-TH 9912123;%%.

\bibitem{Aganagic:2000gs}
M.~Aganagic and C.~Vafa, ``{M}irror symmetry, {D}-branes and counting
  holomorphic discs,''
\href{http://www.arXiv.org/abs/hep-th/0012041}{{\tt hep-th/0012041}}.
%%CITATION = HEP-TH 0012041;%%.

\bibitem{Cheung:2002yw}
Y.-K.~E. Cheung, Y.~Oz, and Z.~Yin, ``Families of {$\N=2$} strings,''
\href{http://arXiv.org/abs/hep-th/0211147}{{\tt hep-th/0211147}}.
%%CITATION = HEP-TH 0211147;%%.

\bibitem{Hori:2000kt}
K.~Hori and C.~Vafa, ``Mirror symmetry,''
\href{http://www.arXiv.org/abs/hep-th/0002222}{{\tt hep-th/0002222}}.
%%CITATION = HEP-TH 0002222;%%.

\bibitem{Penrose:1976js}
R.~Penrose, ``Nonlinear gravitons and curved twistor theory,'' {\em Gen. Rel.
  Grav.} {\bf 7} (1976)
31--52.
%%CITATION = GRGVA,7,31;%%.

\end{thebibliography}

\providecommand{\href}[2]{#2}\begingroup\raggedright\endgroup

\end{document}